\documentclass[preprint,showpacs,preprintnumbers,amsmath,amssymb,nofootinbib]{revtex4}

\usepackage{graphicx}
\usepackage{dcolumn}
\usepackage{bm}

\begin{document}

\title{Capture of slow antiprotons by helium atoms}

\author{J. R\'{e}vai}
\email{revai@rmki.kfki.hu}
 \affiliation{Research Institute for
Particle and Nuclear Physics, H-1525 Budapest, P.O.B. 49, Hungary}

\author{N. V. Shevchenko}
 \email{shev@thsun1.jinr.ru}
\affiliation{Joint Institute  for Nuclear Research, Dubna, 141980,
Russia}

\date{\today}

\begin{abstract}
A consistent quantum mechanical calculation of partial cross sections leading to different
final states of antiprotonic helium atom was performed. For the four-body scattering wave
function, corresponding to the initial state, as well as for the antiprotonic helium wave
function, appearing in the final state, adiabatic approximations were used. Further, symmetric
and non-symmetric effective charge (SEC, NEC) approximations were introduced for the two-electron
wave functions in the field of the two fixed charges of the $He$ nucleus and the antiproton.
Calculations were carried out for a wide range of antiprotonic helium states and incident energies
of the antiproton below the first ionization threshold of the $He$ atom. The origin of the rich
low-energy structure of certain cross sections is discussed in detail.
\end{abstract}

\pacs{36.10.-k, 25.43.+t, 34.90.+q}
\maketitle

%
%
%

\section{Introduction}
One of the most impressive success stories of the last decade in
few-body physics is the high precision experimental and
theoretical studies of long lived states in antiprotonic helium
(for an overview see~\cite{yamazaki}). While the energy levels
have been both measured and
calculated to an extreme precision, allowing even for improvement
of numerical values of fundamental physical constants, some other
relevant properties of these states were studied with considerably
less accuracy. Among these is the formation probability of
different metastable states, characterized by total angular momentum
$J$ and ''vibrational'' quantum number $v$, in the capture reaction
\begin{equation}
\label{react}
\bar p \, + \, ^4He \longrightarrow (^4He^+ \, \bar p)_{Jv} + e^-.
\end{equation}
The existing calculations of the capture rates of slow antiprotons in
$He$~\cite{koren1,koren2,cohen} are based on classical or semiclassical
approaches and they mainly
address the reproduction of the overall fraction (3\%)
of delayed annihilation events. Recent experimental results from
the ASACUSA project~\cite{ASACUSA}, however, yield some information on individual
populations of different metastable states, and our aim is to
perform a fully quantum mechanical calculation of the formation
probability of different states in the capture reaction.

The exact solution of the quantum mechanical four-body problem,
underlying the reaction~(\ref{react}) is far beyond the scope of this work,
and probably also of presently available calculational
possibilities. Still, we want to make a full quantum mechanical,
though approximate, calculation of the above process. Full is
meant in the sense that all degrees of freedom are taken
explicitly into account, all the wave functions we use, are true
four-body states.

\section{Calculation Method}

The partial cross section, leading to a specified final state
$(J,v)$ of the antiprotonic helium can be written as
\begin{equation}
\label{sigmaint}
\sigma_{Jv} = 2 \, (2\pi)^4 \,\frac{K_f}{K_i}\, \mu_i \,\mu_f\int
d\Omega_{{\bf K}_f} \left|\langle\Phi^f_{Jv, {\bf K}_f}|V_f|\,
\Psi^i_{He, {\bf K}_i}\rangle \right|^{\, 2}
\end{equation}
where $\Psi^i_{He, {\bf K}_i}$ is the exact 4-body scattering wave
function corresponding to the initial state
\begin{equation}
\label{phiI}
\Phi^i_{He,\, {\bf K}_i} ({\bf r}_1,{\bf r}_2,{\bf R}) =
\Phi_{He}({\bf r}_1,{\bf r}_2) \,
 \frac{1}{(2 \pi)^{3/2}} \, e^{i {\bf K}_i {\bf R}}_{\qquad,}
\end{equation}
while the final state $\Phi^f_{Jv,\, {\bf K}_f}$ is taken in the
form
\begin{equation}
\label{phiF}
\Phi^f_{Jv,\, {\bf K}_f}
(\mbox{\boldmath{$\rho$}}_1,\mbox{\boldmath{$\rho$}}_2,{\bf R}) =
 \Phi_{Jv}(\mbox{\boldmath{$\rho$}}_1,{\bf R}) \,
 \frac{1}{(2 \pi)^{3/2}} \, e^{i {\bf K}_f
 \mbox{\boldmath{$\rho$}}_2}_{\qquad \;\;.}
\end{equation}
Here ${\bf r}_i$ are the vectors pointing from the helium nucleus to the
$i$-th electron, ${\bf R}$ is the vector between $He$ and $\bar p$,
and $\mbox{\boldmath{$\rho$}}_i$ are the Jacobian coordinates of the
electrons, measured from the $He-\bar p$ center of mass:
\begin{equation}
 {\bf r}_i = \mbox{\boldmath{$\rho$}}_i + \alpha \, {\bf R};
 \qquad \alpha = \frac{m_{\bar p}}{m_{\bar p}+m_{He}} \,,
\end{equation}
while $\mu_i$ and $\mu_f$ are the reduced masses in initial and final
channels, respectively.
In Eq.~(\ref{phiI}) $\Phi_{He}({\bf r}_1,{\bf r}_2)$ denotes the $He$ ground
state wave function, while in Eq.~(\ref{phiF})
$\Phi_{Jv}(\mbox{\boldmath{$\rho$}}_1,{\bf R})$ is the antiprotonic
helium final state, for which we used a Born-Oppenheimer form~\cite{shim,revai}:
\begin{equation}
\label{wfpHe}
 \Phi_{Jv}(\mbox{\boldmath{$\rho$}},{\bf R}) =
 \frac{\chi_{Jv}(R)}{R}\,
Y_{JM}(\hat{R}) \; \varphi_{ 1\sigma}^{(2,-1)}(\mbox{\boldmath{$\rho$}};{\bf R})
\end{equation}
where $\varphi_{ 1\sigma}^{ (Z_1,Z_2)}(\mbox{\boldmath{$\rho$}};{\bf R})$ is a
two-center wave function, describing the electron (ground state)
motion in the field of two charges $(Z_1,Z_2)$, separated by a fixed
distance $R$:
\begin{equation}
\left(-\frac{1}{2} \Delta_{\displaystyle \bf{r}} + \frac{Z_1}{r}
+ \frac{Z_2}{|{\bf r} - {\bf R}|} \right)
\; \varphi_{ 1\sigma}^{ (Z_1,Z_2)}({\bf r};{\bf R}) \; = \;
\varepsilon_{1\sigma}^{(Z_1,Z_2)}(R) \; \varphi_{1\sigma}^{ (Z_1,Z_2)}({\bf r};{\bf R})
\end{equation}
while $\chi_{Jv}(R)$ is the heavy-particle relative
motion wave function, corresponding to $({}^4He \, \bar{p} \; e^-)$
angular momentum $J$ and ''vibrational'' quantum number $v$:
\begin{equation}
\label{khiatomc}
\left(-\frac{1}{2 \mu} \left[ \frac{d^2}{dR^2} - \frac{J(J+1)}{R^2} \right]
 - \frac{2}{R}\;+ \; \varepsilon_{ 1\sigma}^{(2,-1)}(R) - E_{J,v} \right)
\; \chi_{Jv}(R) = 0 ,\quad
\footnote{Since in the following we shall deal only with $1\sigma$ two-center
ground states, the index $1\sigma$ will be omitted throughout the paper.}
\end{equation}
$\mu$ being the $He - \bar p$ reduced mass.

The transition potential in Eq.~(\ref{sigmaint}) is obviously the interaction of
the emitted electron (\#2) with the rest of the system:
\begin{equation}
\label{3pots}
 V_f = - \, \frac{2}{{\bf r}_2} + \frac{1}{|{\bf r}_2 - {\bf R}|} +
  \frac{1}{|{\bf r}_1 - {\bf r}_2|}_{\, .}
\end{equation}
The electron anti-symmetrization is accounted for by taking an
$r_1 \Longleftrightarrow r_2$ symmetric initial state wave function $(S=0)$
and the factor 2 in front of the cross-section~(\ref{sigmaint}), reflecting the
indistinguishability of emitted particles \cite{ekstein}. \\

The general expression~(\ref{sigmaint}) for the cross-section, leading to a specific state
$(J,v)$ can be rewritten in terms of matrix elements between angular momentum
eigenstates as
\begin{equation}
\label{sigmasum}
 \sigma_{Jv} = 2\,(2\pi)^4 \,\frac{K_f}{K_i}\, \mu_i \,\mu_f
 \sum_{J_t,l} (2 J_t + 1) \,
 |M_{J,l}^{J_t}|^2
\end{equation}
with
\begin{equation}
\label{me}
 M_{J,l}^{J_t}=
 \langle \, [\Phi_{Jv}(\mbox{\boldmath{$\rho$}}_1,{\bf R})
 \, \phi_{K_f,l}(\mbox{\boldmath{$\rho$}}_2)]_{M_t}^{J_t} \,  | \,
 V_f \, | \, \Psi_{He,K_i}^{i \, J_t,M_t}(\mbox{\boldmath{$\rho$}}_1,
 \mbox{\boldmath{$\rho$}}_2,{\bf R}) \, \rangle,
\end{equation}
where $[\,\,\,]_M^J$ stands for vector coupling,
$\Psi_{He,K_i}^{i \, J_t,M_t}$ is the exact scattering wave
function with total angular momentum $J_t$, corresponding to
the initial state
$$
[\Phi_{He}^{J=0}({\bf r}_1,{\bf r}_2)
 \, \phi_{K_i,J_t}({\bf R})]_{M_t}^{J_t}
$$
and $\phi_{K_i,l}({\bf r})$ denotes free states with definite
angular momentum
$$
 \phi_{K,l}({\bf r}) = \sqrt{\frac{2}{\pi}} \, j_l(Kr) Y_{lm}(\hat{r}) .
$$
It can be seen from Eqs.~(\ref{sigmasum},\ref{me}), that a given antiprotonic
helium final state $(J,v)$ can be formed from different total
angular momentum states, depending on
the orbital momentum $l$, carried away by the emitted electron. \\

The simplest way of approximate evaluation of Eq.~(\ref{sigmaint}) or~(\ref{sigmasum})
is to use Born approximation, replacing the exact scattering wave function
$\Psi^i_{He,{\bf K}_i}$ by its asymptotic form $\Phi^i_{He,{\bf K}_i}$
from Eq.~(\ref{phiI}). In order to get an idea of the feasibility of such a
''full'' (including all degrees of freedom)
calculation we evaluated the cross-sections $\sigma_{Jv}$
in Born approximation in a wide range of quantum numbers $(J,v)$. For the
$He$ ground state wave function in this case we used the simplest
variational form
\begin{equation}
\label{wfHe}
 \Phi_{He}(\bf{r}_1,\bf{r}_2) = N \, \exp{(-\sigma (r_1 + r_2))}
\end{equation}
with $\sigma=27/16$ taken from book~\cite{bethe}. In spite of the
known poor quality of the Born approximation for slow collisions,
due to the realistic final state wave function, we hoped to get some
information at least about the relative population probabilities of
different final states. These expectations were not confirmed, the
Born cross-sections turned to be orders of magnitude away from the
more realistic ones. The detailed results of the Born calculation
can be found in~\cite{ourBorn}.

There are two basic drawbacks of the Born approximation for slow collisions
and long-range forces:

 --- the antiproton ''feels'' the interaction from the $He$ atom, it approaches,
 therefore, its wave function in the form of a plane wave has to be modified,

 --- the $He$ electrons also ''feel'' the approaching antiproton, the polarization of
 their wave functions has to be taken into account.

To meet these requirements we used an adiabatic, Born-Oppenheimer type
approximation for the wave function $\Psi^i$:
\begin{equation}
\label{psiI}
\Psi_{He,{\bf K}_i}^i \approx
\Phi_{He}({\bf r}_1,{\bf r}_2;{\bf R}) \, \chi_{{\bf K}_i}({\bf R}),
\end{equation}
where $\Phi_{He}({\bf r}_1,{\bf r}_2;{\bf R})$ is the ground state
wave function of the $He$ atom in the presence of a negative unit
charge (the antiproton) at a distance $R$ from the $He$ nucleus:
\begin{equation}
\label{wfHe2centr} \mathcal{H}_{He,\, \bar p}(R) \, \Phi_{He}({\bf
r}_1,{\bf r}_2;{\bf R}) \; = \; \varepsilon(R) \, \Phi_{He}({\bf
r}_1,{\bf r}_2;{\bf R}),
\end{equation}
$$
\mathcal{H}_{He,\, \bar p}(R) =  -\frac{1}{2} \Delta_{\displaystyle
\bf{r}_1} -\frac{1}{2} \Delta_{\displaystyle \bf{r}_2} -
\frac{2}{r_1} - \frac{2}{r_2} + \frac{1}{|{\bf r}_1 - {\bf r}_2|} +
\frac{1}{|{\bf r}_1 - {\bf R}|} + \frac{1}{|{\bf r}_2 - {\bf R}|} ;
$$
and $\chi_{{\bf K}_i}({\bf R})$ is the antiproton scattering wave function
in the adiabatic $He - \bar p$ potential:
\begin{equation}
\label{VHepbar}
V_{He - \bar p}(R) = - \frac{2}{R} + \varepsilon(R).
\end{equation}
\begin{equation}
\left( -\frac{1}{2 \mu} \Delta_{\displaystyle \bf R} + V_{He - \bar
p}(R) \right) \, \chi_{{\bf K}_i}({\bf R}) \; = \; \frac{K_i^2}{2
\mu} \; \chi_{{\bf K}_i}({\bf R}) .
\end{equation}

In this approach the most difficult task is the solution of
Eq.~(\ref{wfHe2centr}), the determination of the wave function of two
interacting electrons in the field of two fixed charges. Instead of
performing a cumbersome variational calculation, as e.g.
in~\cite{ahlrichs,gibbs}, we follow an approximation scheme proposed by
Briggs, Greenland, and Solov'ev (BGS)~\cite{briggs}, according to
which the solution of Eq.~(\ref{wfHe2centr}) can be sought in the form
of two single-electron two-center wave functions:
\begin{equation}
\Phi_{He}({\bf r}_1,{\bf r}_2;{\bf R}) \;
\approx \; \varphi^{(Z_{11},Z_{12})}({\bf r}_1;{\bf R}) \,
\varphi^{(Z_{21},Z_{22})}({\bf r}_2;{\bf R})
\end{equation}
with
\begin{equation}
\left(-\frac{1}{2} \Delta_{\displaystyle \bf{r}} + \frac{Z_{i1}}{r}
+ \frac{Z_{i2}}{|{\bf r} - {\bf R}|} \right)
\; \varphi^{ (Z_{i1},Z_{i2})}({\bf r}_1;{\bf R}) \; = \;
\varepsilon^{(Z_{i1},Z_{i2})}(R) \; \varphi^{ (Z_{i1},Z_{i2})}({\bf r}_1;{\bf R})
\end{equation}
and the $\varepsilon(R)$ of Eqs.~(\ref{wfHe2centr},\ref{VHepbar}) is
\begin{equation}
\label{ERmed}
\varepsilon(R) = \varepsilon^{(Z_{11},Z_{12})}(R) + \varepsilon^{(Z_{21},Z_{22})}(R) .
\end{equation}
In this construction the effect of the electron-electron interaction $|{\bf r}_1-{\bf r}_2|^{-1}$
in Eq.~(\ref{wfHe2centr}) is approximated by suitable choice of the effective charges $(Z_{11},Z_{12},
Z_{21},Z_{22})$.
BGS suggest to fix the effective charges ''seen'' by the first electron,
$Z_{11}$ and $Z_{12}$,  at the real charges of $He$ and $\bar p$,
while those for the second one,
$Z_{21}$ and $Z_{22}$, can be obtained from the requirement, that in the two limiting cases
$R \rightarrow 0$ and $R \rightarrow \infty$, the ground state energies of $H^-$ ion
and $He$ atom should be reproduced:
\begin{equation}
\label{ERnulinf}
\varepsilon(R \rightarrow 0) = E_{gs}(H^-), \qquad
\varepsilon(R \rightarrow \infty) = E_{gs}(He) \, .
\end{equation}
The conditions (\ref{ERnulinf}) are fulfilled for
\begin{eqnarray}
\nonumber
&{}&Z_{11}= 2.0, \qquad \phantom{441} Z_{12}=-1.0, \\
\label{z1}
&{}&Z_{21}= 1.3444, \qquad Z_{22}=-1.1095.
\end{eqnarray}
For intermediate $R$-s $\varepsilon(R)$ is given by~(\ref{ERmed}).

As for $He$ wave function, the two electrons in this case are treated in a non-symmetric
way, and the wave function has to be symmetrized explicitly:
\begin{eqnarray}
\nonumber
\Phi_{He}({\bf r}_1,{\bf r}_2;{\bf R})
=  N(R) \left[
\varphi^{(Z_{11},Z_{12})}({\bf r}_1;{\bf R}) \,
\varphi^{(Z_{21},Z_{22})}({\bf r}_2;{\bf R}) \, + \, \right. \\
\left. + \,
\varphi^{(Z_{11},Z_{12})}({\bf r}_2;{\bf R}) \,
\varphi^{(Z_{21},Z_{22})}({\bf r}_1;{\bf R})
\right] .
\label{phiNEC}
\end{eqnarray}
There is, however, a more symmetric realization of the BGS idea: starting with
the plausible requirement, that the two electrons should ''see'' identical effective
charges: $Z_{11} = Z_{21}$, $Z_{22} = Z_{12}$
we still can impose the conditions~(\ref{ERnulinf}) for $R \rightarrow 0$ and
$R \rightarrow \infty$, only in this case the $\varepsilon(R)$ will be
the sum of two equal single-particle energies:
$$
\varepsilon(R) = 2 \, \varepsilon^{(Z_{11},Z_{22})}(R).
$$
For this case we get
\begin{equation}
\label{z2}
Z_{11}= 1.704, \qquad Z_{22}=-0.9776.
\end{equation}

The $\varepsilon(R)$ in this case is very similar to the previous
one, maybe a little closer to the ''quasi-exact'' variational curve. In this
second case --- for brevity let us call it SEC (Symmetric Effective Charge), in
contrast to the NEC (Non-symmetric Effective Charge) case --- the wave function
is simply
\begin{equation}
\label{phiSEC}
\Phi_{He}({\bf r}_1,{\bf r}_2;{\bf R}) \;
= \; \varphi^{(Z_{11},Z_{22})}({\bf r}_1;{\bf R}) \,
\varphi^{(Z_{11},Z_{22})}({\bf r}_2;{\bf R}) \, .
\end{equation}
\begin{figure}
\includegraphics[scale=0.4,angle=-90]{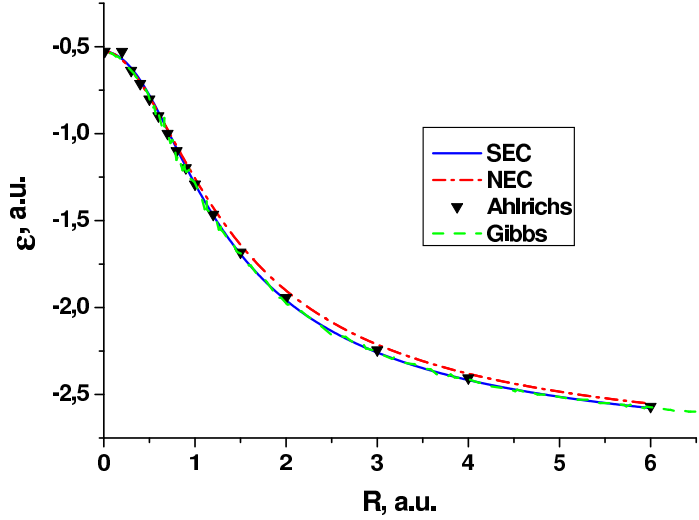}
\caption{\label{genius.fig} Electronic energies $\varepsilon(R)$ for the NEC, SEC and variational
cases.}
\end{figure}
The differences between electronic  energies $\varepsilon(R)$ for the NEC,
SEC and variational calculations (performed by Ahlrichs {\it et al}~\cite{ahlrichs}
and Gibbs~\cite{gibbs}) are shown on Fig.~\ref{genius.fig}. It is seen that
both cases reproduces the variational results remarkably well, while SEC is practically
indistinguishable from the more recent of them~\cite{gibbs}. \\

For both choices~(\ref{phiNEC}) and~(\ref{phiSEC}) the definite total angular momentum wave function
corresponding to~(\ref{psiI}) can be written as
\begin{equation}
\label{psiIfull}
\Psi^{iJ_tM_t}_{He,K_i}({\bf r_1},{\bf r_2},{\bf R})=\Phi_{He}({\bf r_1},{\bf r_2};{\bf R})
\frac{\chi^{J_t}_{K_i}(R)}{K_iR}Y_{J_t,M_t}(\hat R)\; ,
\end{equation}
since the $1\sigma$ ground state two-center functions $\varphi({\bf r};{\bf R})$ do not carry
any total angular momentum: they are eigenfunctions of $\hat J^2=(\hat l_{\bf r}+\hat L_{
\bf R})^2$ with zero eigenvalue, even if they are not eigenfunctions of $\hat l_{\bf r}^2$
and $\hat L_{\bf R}^2$ separately. The function $\chi^{J_t}_{K_i}(R)$ satisfies the equation
\begin{equation}
\label{khiKi}
\left[ \frac{d^2}{dR^2} -2\mu{\mathsf V}_{\!J_{\,t}}(R)
 + {K_i^2} \right] \, \chi^{J_t}_{K_i}(R) \; = \; 0
\end{equation}
with the effective $He - \bar p$ potential
\begin{equation}
\label{efpot}
\mathsf V_{\!J_{\,t}}(R)=\frac{J_t(J_t+1)}{2\mu R^2}+ V_{He - \bar p}(R)\;.
\end{equation}
To solve Eq.~(\ref{khiKi}) numerically, first, the asymptotic form of $\chi^{J_t}_{K_i}(R)$ has to be
clarified. The asymptotic behavior of the $1\sigma$ two-center energies can be written as
\begin{equation}
\varepsilon^{(Z_1,Z_2)}(R)\mathop{\longrightarrow}\limits_{R \to \infty} - \frac{Z_1^2}{2}
- \frac{Z_2}{R} + \mathcal O(R^{-4})
\end{equation}
and thus using Eqs.~(\ref{VHepbar}) and~(\ref{ERmed}) we get
\begin{equation}
\label{VRinf}
V_{He - \bar p}(R)\mathop{\longrightarrow}\limits_{R \to \infty} - \frac{Z_{11}^2+
Z_{21}^2}{2}- \frac{2+Z_{12}+Z_{22}}{R} + \mathcal O(R^{-4})
\end{equation}
Dropping the irrelevant constant term from~(\ref{VRinf}) we see, that asymptotically it corresponds to
a Coulomb-interaction with effective charge $Z_{as}=-(2+Z_{12}+Z_{22})$. From the actual values
of $Z_{12}$ and $Z_{22}$~(\ref{z1}) and~(\ref{z2}) we can conclude, that NEC corresponds to a weak repulsion,
while SEC --- to an even weaker attraction. In reality, of course, there is no $1/R$ term in
the asymptotic $He - \bar p$ interaction, since the $He$ atom is neutral.

\begin{figure}
\includegraphics[scale=0.65,angle=-90]{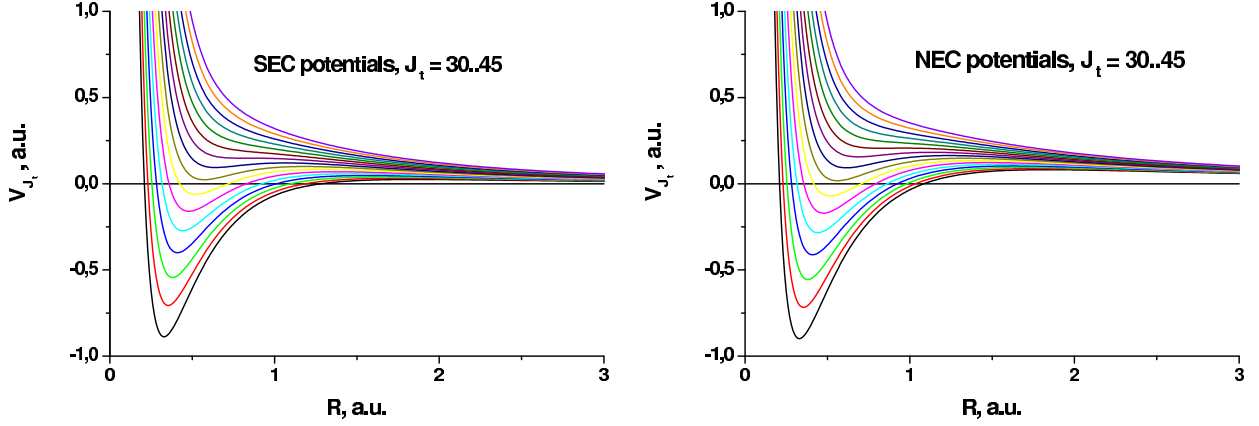}
\caption{\label{potentials.fig} Effective $He - \bar p$ potentials
$\mathsf V_{\!J_{\,t}}(R)$ for different $J_t$ values.}
\end{figure}

It has to be noted, that in spite of the closeness of the NEC and SEC electron energies on
Fig.~\ref{genius.fig},
when we include the centrifugal term, the depth of the minima and the height of the potential
barriers differ significantly (see Fig.~\ref{potentials.fig}) and this fact strongly influences the low
energy capture cross sections.

According to~(\ref{VRinf}) Eq.~(\ref{khiKi}) has to be solved with the asymptotic condition
\begin{equation}
\label{khiRinf}
\chi^{J_t}_{K_i}(R)\mathop{\longrightarrow}\limits_{R \to \infty} \cos \delta_{J_t}(K_i)
F_{J_t}(\eta,K_iR)+\sin \delta_{J_t}(K_i)G_{J_t}(\eta,K_iR)\;,
\end{equation}
where $F_{J_t}$ and $G_{J_t}$ are the regular and irregular Coulomb wave functions, with
Sommerfeld-parameter
\begin{equation}
\eta=\frac{Z_{as}\mu}{K_i}
\end{equation}
and $\delta_{J_t}(K_i)$ is the phase shift caused by the non-coulombic part of the
potential. After the numerical solution of Eq.~(\ref{khiKi}) with boundary
conditions~(\ref{khiRinf}) the matrix elements~(\ref{me})
entering the formula~(\ref{sigmasum}) for the cross sections can be calculated by numerical
integration.

\section{Results and Discussion}

\begin{figure}
\includegraphics[scale=0.7,angle=0]{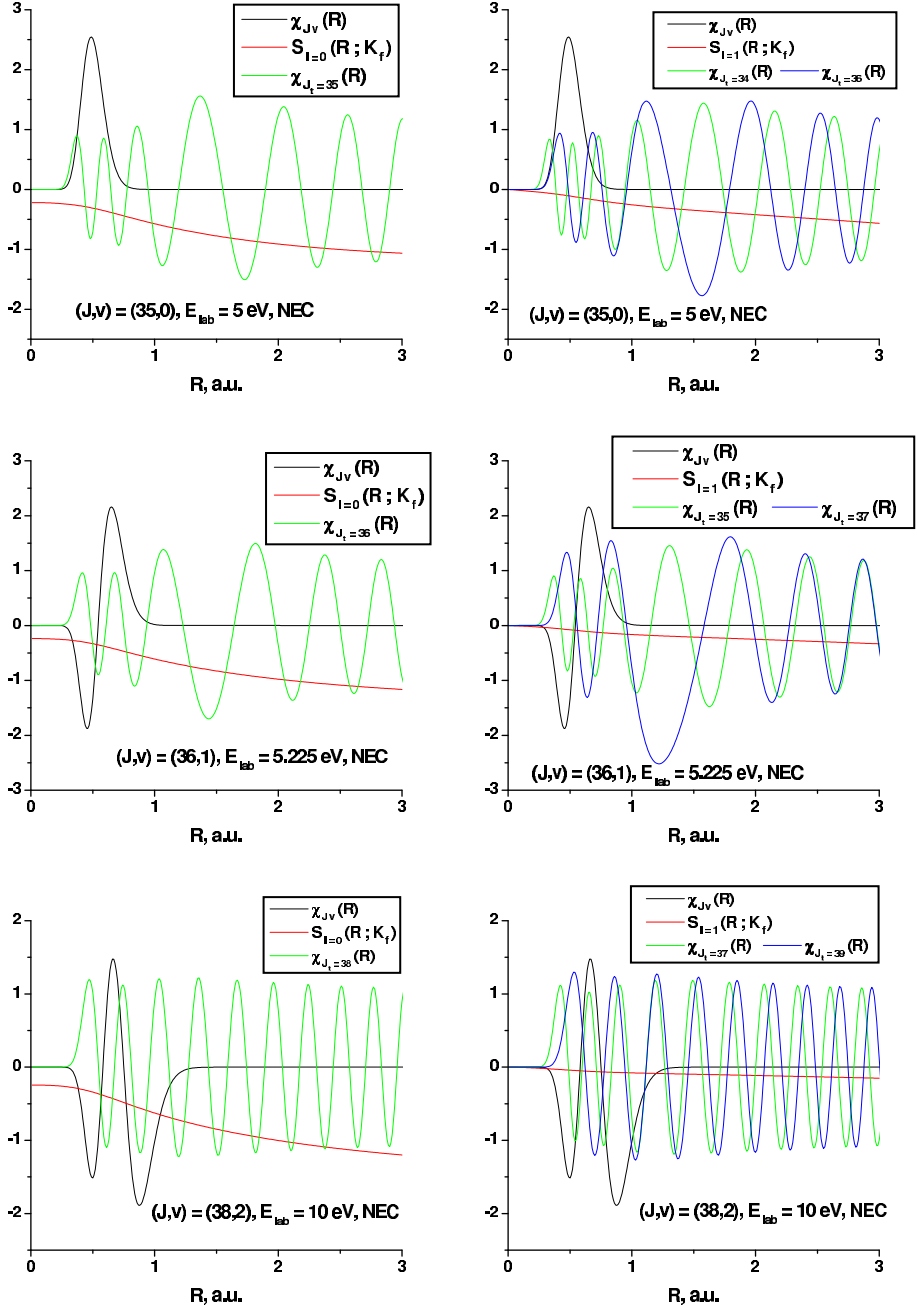}
\caption{\label{FJv.fig} Functions $\chi_{Jv}$, $\chi_{K_i}^{J_t}$ and $S_l$ for different sets of
parameters. }
\end{figure}
We start the discussion of our results by the remark, that the expression~(\ref{me}) for the matrix
element $M^{J_t}_{J,l}$ in our adiabatic approximation can be rewritten as
\begin{equation}
\label{Mint}
M^{J_t}_{J,l} \sim \int \chi_{Jv}(R)S_l(R;K_f)\chi^{J_t}_{K_i}(R) dR \;,
\end{equation}
where $\chi_{Jv}(R)$ and $\chi^{J_t}_{K_i}(R)$ are the $He - \bar p$ relative motion wave
functions, introduced in Eqs.~(\ref{khiatomc}) and~(\ref{psiIfull}),
respectively, while $S_l(R;K_f)$ contains ''all
the rest'': the three potentials~(\ref{3pots}) integrated over electron wave functions and coordinates,
angular variables of {\bf R} and summations over intermediate quantum numbers. This representation
is useful, since it turns out, that the basic dependence of the matrix elements on the quantum
numbers and incident energy is contained in the two $\chi$ functions, while $S_l(R;K_f)$ weakly
and smoothly depends on its arguments with a significant decrease with increasing $l$ --- the
orbital momentum of the emitted electron. For a few selected cases the three functions in the
integrand of Eq.~(\ref{Mint}) are shown in Fig.~\ref{FJv.fig}.
This feature of $S_l(R;K_f)$ allows the interpretation of Eq.~(\ref{Mint}) as a
matrix element of antiproton transition from the initial state $\chi^{J_t}_{K_i}(R)$ into a
final state $\chi_{Jv}(R)$ under the action of the effective potential $S_l(R;K_f)$.

\begin{figure}
\includegraphics[scale=0.7,angle=0]{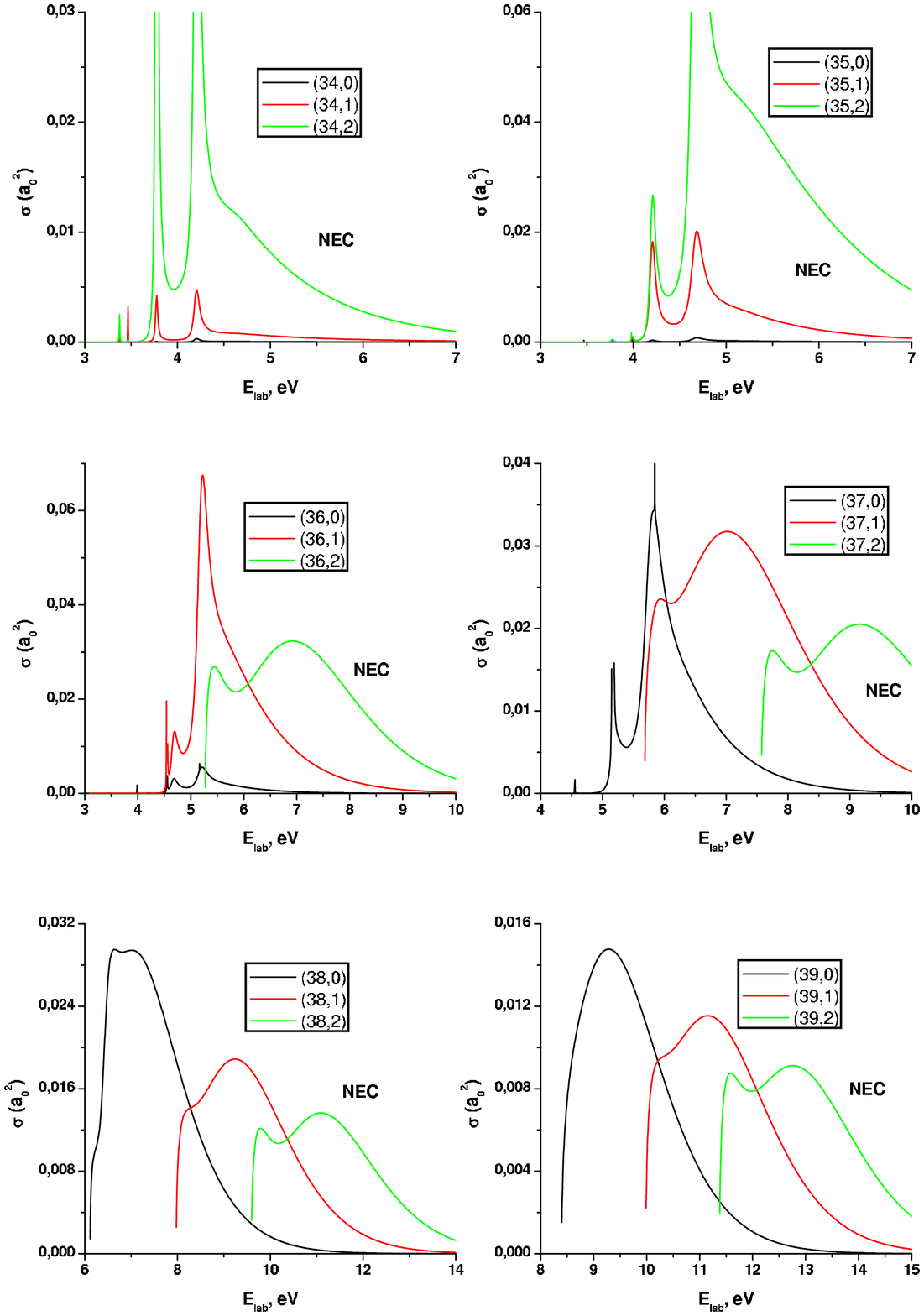}
\caption{\label{diffEsNEC.fig} Energy dependence of the cross-sections for
some $(J,v)$ states, NEC case.}
\end{figure}

\begin{figure}
\includegraphics[scale=0.7,angle=0]{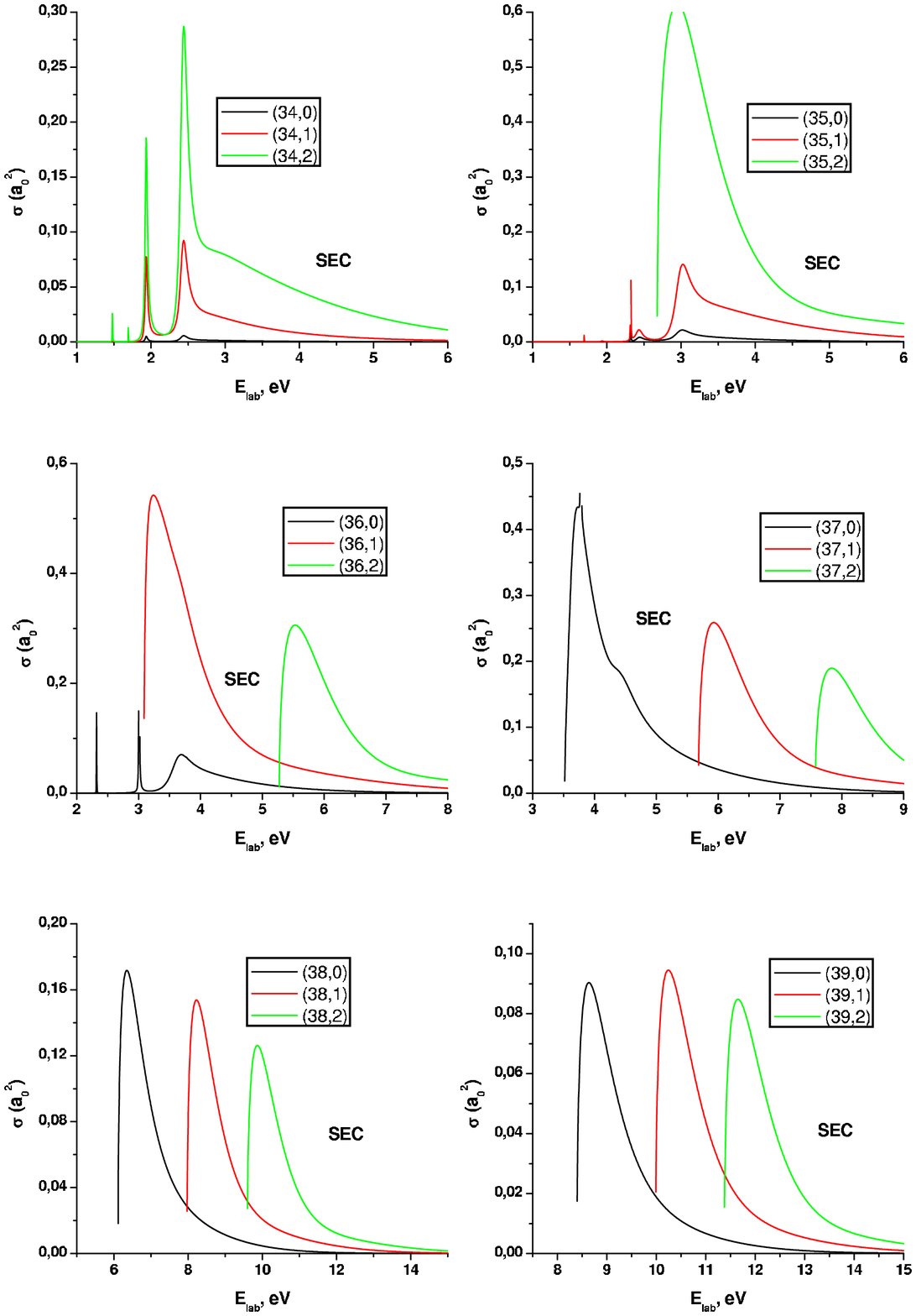}
\caption{\label{diffEsSEC.fig} Energy dependence of the cross-sections for
some $(J,v)$ states, SEC case.}
\end{figure}

We have calculated the capture cross sections leading to different final states for antiproton
energies below the first ionization threshold $E_{lab}=30.8\; eV$.
The overall energy dependence of the NEC and SEC cross sections $\sigma_{Jv}(E)$ is shown in
Figs.~\ref{diffEsNEC.fig},~\ref{diffEsSEC.fig}
for a few quantum numbers from the region of expected largest capture probability.
All cross-sections are measured in units of $a_0^2$, $a_0$ being the atomic length unit.
The main features of these curves can be summarized as follows.

Obviously, final states with
energy below the $He$ atom ground state energy (-2.9036 a.u.) have a positive $Q$ value, so
they can be reached for arbitrary low antiproton energy. For example, on
Figs.~\ref{diffEsNEC.fig},~\ref{diffEsSEC.fig}
there are such states: with
$(J=34, v=0,1,2),(J=35,v=0,1)$, and $(J=36,v=0)$. States with higher energy can be excited only
above their threshold energies; the steep rise of the cross sections above these thresholds
can be clearly seen.

\begin{figure}
\includegraphics[scale=0.65,angle=-90]{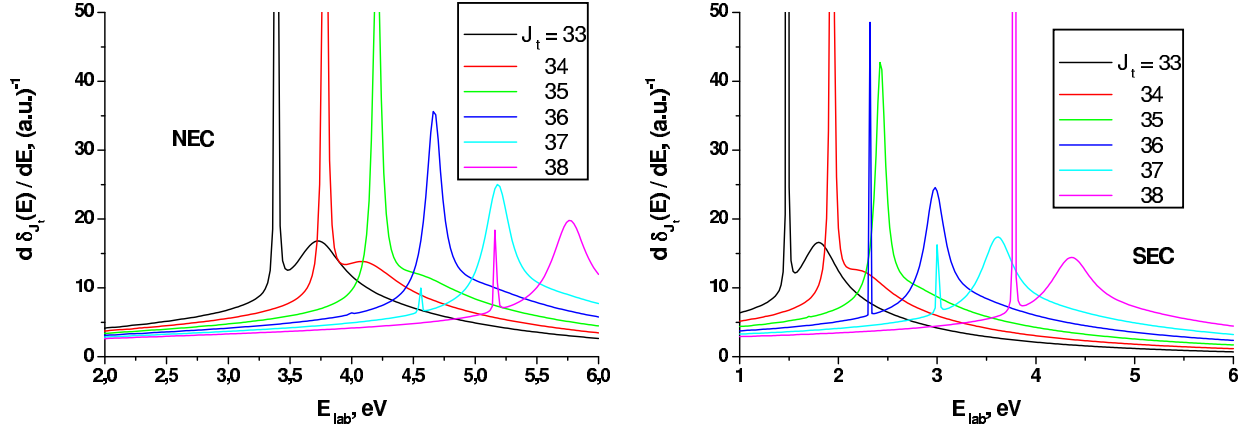}
\caption{\label{difdels.fig} Time delays for different values of total angular momentum $J_t$,
NEC and SEC cases.}
\end{figure}

Another remarkable feature of certain cross sections is their rich low energy structure. This
is due to the repulsive barriers of the effective potentials $\mathsf V_{\!J_{\,t}}$ for
$J_t\leq 38-39$, as seen on Fig.~\ref{potentials.fig}.
These barriers, in general, strongly suppress the penetration of $\chi^{J_t}_{K_i}(R)$
into the interior region, thus reducing the sub-barrier cross sections. For certain sub-barrier
energies, however, there are quasi-stationary states in these potentials, when the interior wave
function has a large amplitude, leading to sharp resonances in the cross sections. In order to
clarify the origin of these peaks, we looked at the energy dependence of the phase shifts
$\delta_{J_t}$ of Eq.~(\ref{khiRinf}). On Fig.~\ref{difdels.fig}
we plotted the quantity $d\delta_{J_t}(E)/dE$ --- the so
called time delay --- which for isolated resonances is very similar to the more familiar
Breit-Wigner  cross section curve. It can be seen, that for all angular momenta $J_t$
for which the potential has a barrier, there is a narrow resonance which is correlated with a
corresponding peak in the capture cross section. A given cross section curve may contain several
of these peaks, corresponding to different $J_t$ and $l$ values contributing to formation of
a given final state, according to the sum in Eq.~(\ref{sigmasum}). In general, it is interesting
to note, that in contrast to a common belief, the sum of Eq.~(\ref{sigmasum}) is not dominated
by the $s$-electron emission $(J=J_t,l=0)$ term, the $p$-electrons practically always, while
the $d$-electrons in certain cases contribute significantly. The reason for this may be, that
the decrease of $S_l(R;K_f)$ with growing $l$ could be ''compensated'' by the possibility of lower
$J_t$ values, for which the effective potentials contain less repulsion, thus allowing more
penetration of $\chi_{K_i}^{J_t}$ into the interior region.

The $d\delta_{J_t}(E)/dE$ plots apart from the narrow peaks corresponding to quasi-stationary
states, show another, much broader peak, in some cases superposed on the narrow one. This one is
connected with specific behavior of elastic scattering when the energy is close to the
potential maximum; it is called ''orbiting''~\cite{newton}.

\begin{figure}
\includegraphics[scale=0.6,angle=-90]{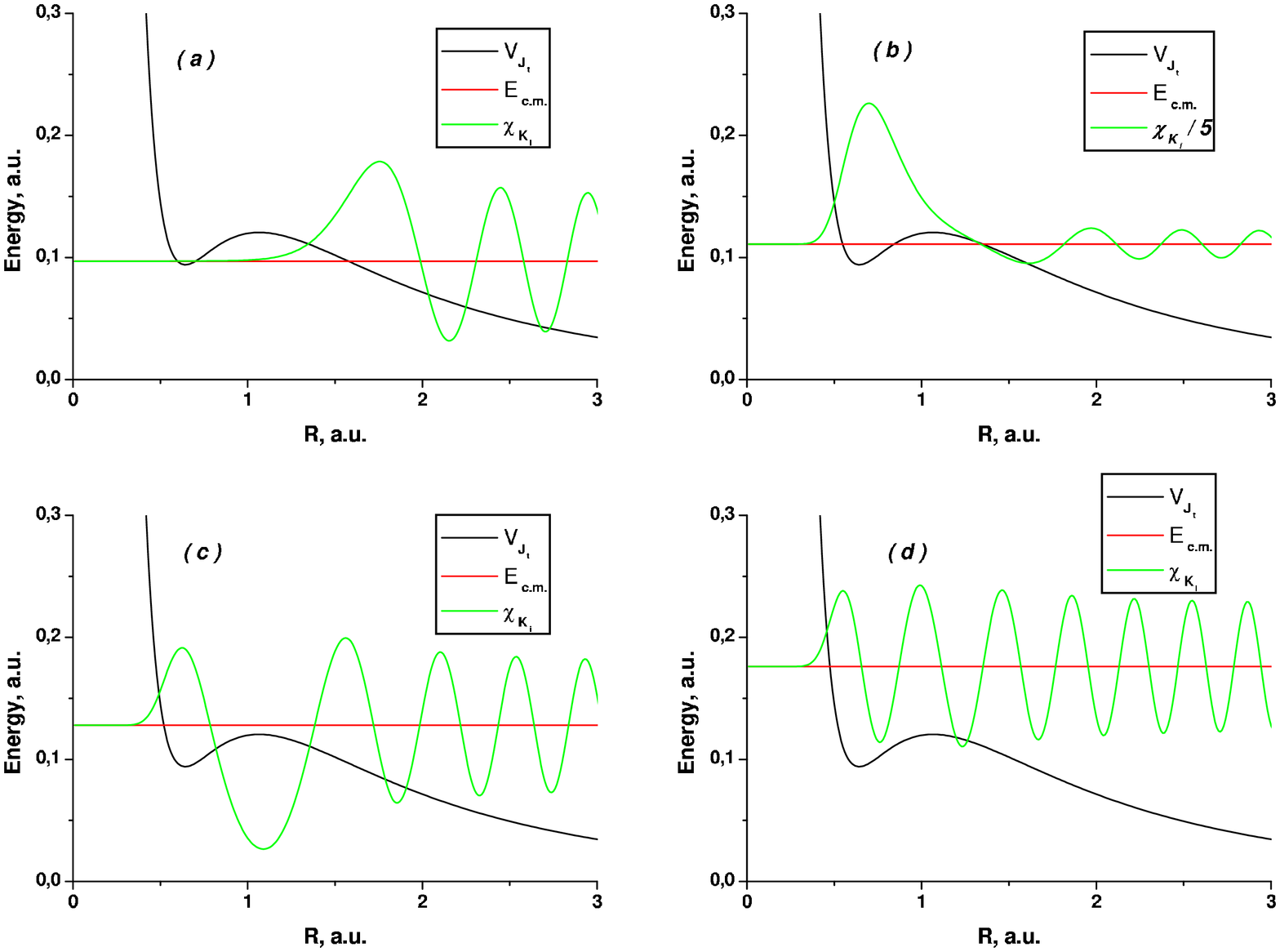}
\caption{\label{khiev.fig}
Incident antiproton wave functions (in arbitrary units) with their potentials. The incident
energies (in CM, the baselines for the wave function plots) are chosen to represent different
cases with respect to barrier maxima.\newline
a) -- sub-barrier non-resonant energy; b) -- sub-barrier resonant energy; c) -- ''orbiting'':
energy close to the potential maximum; d) -- above-barrier energy. }
\end{figure}

The behavior of the incident antiproton wave function $\chi_{K_i}(R)$ for different energy-regimes
with respect to the barrier maximum are illustrated in Fig.~\ref{khiev.fig}.

Final states with higher $J$, for which the relevant effective potentials have no barrier
show a simple energy dependence: a steep rise above the threshold and then an exponential decay
for higher energies. The exponential fall of the cross sections for increasing energies
is characteristic for both barrier-posessing and barrier-less potentials and is due to
increasingly rapid oscillations of $\chi^{J_t}_{K_i}(R)$ in the interior region which reduce the
value of the integral in Eq.~(\ref{Mint}).

\begin{figure}
\includegraphics[scale=0.55,angle=-90]{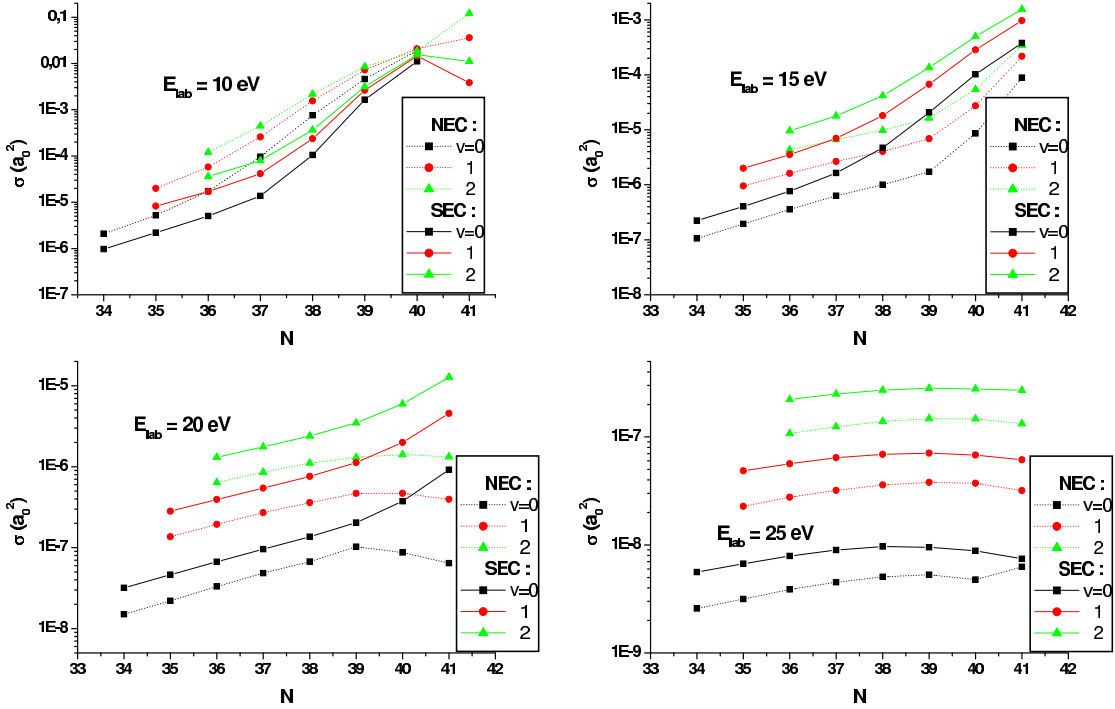}
\caption{\label{vconst.fig} Cross-sections for the lowest few
vibrational quantum numbers $v$ and different incident antiproton
energies (NEC and SEC calculations) plotted against the principal
quantum number $N=J+v+1$.}
\end{figure}

The quantum number dependence of certain cross sections is shown in Fig.~\ref{vconst.fig}
for some above-barrier energies, where such a comparison makes sense.

\section{Conclusions}

To our knowledge, this is the first fully quantum mechanical calculation of the process~(\ref{react}),
with all degrees of freedom taken explicitly into account. The adiabatic wave functions used both for
initial and final states seem to be reasonably realistic. The results show, that quantum mechanical
treatment is really necessary, especially in the low-energy region, where barrier penetration
and resonance effects are essential. The energy dependence of the calculated cross sections
show, that the different final states $(J,v)$ are excited with a large probability in a relatively
narrow window of the incident antiproton energy. In principle, this property could be used for
selective excitation of certain states. On the other hand, the strong energy dependence of the
cross sections prevents us from making statements about the experimentally observable population
numbers of different states since the initial energy distribution of the antiprotons before the
capture is unknown. Even if this distribution was known, the observed and calculated population
numbers could deviate due to collisional (or other) de-excitation of states in the time interval
between the capture and the measurement. Nevertheless, we plan to make calculation of primary
populations taking some trial energy distributions for the antiprotons.

In the discussion of our results we deliberately did not take a
stand concerning the NEC and SEC approximations. In general, the
structure of both cross sections (energy- and quantum number
dependence) is similar, however, SEC gives considerably larger
cross sections, probably due to the somewhat larger attraction of
the SEC effective potentials.  We personally think, that SEC is
physically more realistic,
and the coincidence of SEC's electronic energies with those of recent
variational calculation~\cite{gibbs} can be seen as some confirmation for
this point of view.

\begin{acknowledgments}
One of the authors (JR) acknowledges the support from OTKA grants
T037991 and T042671, while (NVS) is grateful for the hospitality
extended to her in the Research Institute for Particle and Nuclear
Physics, where a significant part of the work has been done. The authors wish to
thank A.T. Kruppa  for providing them with one of the necessary
computer codes.
\end{acknowledgments}

\bibliography{article}

\end{document}